# Wind Speed Data Analysis for Various Seasons during a Decade by Wavelet and S transform

Sabyasachi Mukhopadhyay[1], Prasanta K. Panigrahi[1]
[1]Indian Institute of Science Education and Research Kolkata, Mohanpur 741246, India

***ABSTRACT***

*The appropriate weather prediction is a challenging task and it can be feasible with proper wind speed fluctuation analysis. In this current paper daubechies-4 wavelet is used to analyze the winter wind speed fluctuations due to lesser agitated wind data samples of winter. In summer abrupt changes in wind speed occurs which creates difficulty for wavelets to keep proper track of wind speed fluctuations. So, in that case the concept of the S-transform is introduced.*

## 1. Introduction

The present paper is focused on time-frequency analysis of wind speed data. Currently wind energy is a hot topic as it is one of the important non-conventional energies. We know that Weibull distribution is a very popular tool for wind energy purpose [2, 5]. The usefulness of signal processing tools are increasing in case of wind engineering purpose. For low wind speed analysis purpose use of discrete Hilbert transform (DHT) along with weibull distribution is shown in ref [2]. The discrete Hilbert Transform (DHT) can be used as minimum phase type filter for characterizing and forecasting purpose of Wind speed data is showed by Mukhopadhyay et al., [4]. Mukhopadhyay et al., also showed the optimized DHT and RBF neural network basis analysis for Wind power forecasting purpose [3]. The utility of Wavelet in weather related application was already shown in Ref [1, 13-14]. The application of wavelet transform in the field of Ocean technology was shown by Liu et al., [8]. In 2005, the wavelet transform was used for wind data simulation of Saudi Arabia region by Siddiqi et al., [10]. Thereafter in 2010, the wavelet transform was used to analyze the wind data in Dongting lake cable stayed bridge region by He et al., [9]. Wind speed data of summer of Eastern region of India was analysed by the continuous wavelet transform and multifractality by Mukhopadhyay et al., [7]. Mukhopadhyay et al., already demonstrated the efficacy of S-transform for precancer detection and weather forecasting in their earlier works [12, 15]. In this current work, authors are basically interested on the analysis of wind speed fluctuations over a decade on the basis of Wavelet and S-transform.

## 2. Methodology

The methodology for wind speed fluctuations analysis by S-transform is shown in detail below. In the current work, the below flow chart is followed:

Wind data ⟶ Median Filtering ⟶ S-Transform ⟶ The desired output

**Figure-1** Flow chart of steps for s-transform analysis of Wind data fluctuation of summer

## 3. Results and Discussions

Daubechies-4 Wavelets are useful for wind speed fluctuation analysis of a decade consists of 3months, such as- December, January and February, shown in Figure-2a-2c.

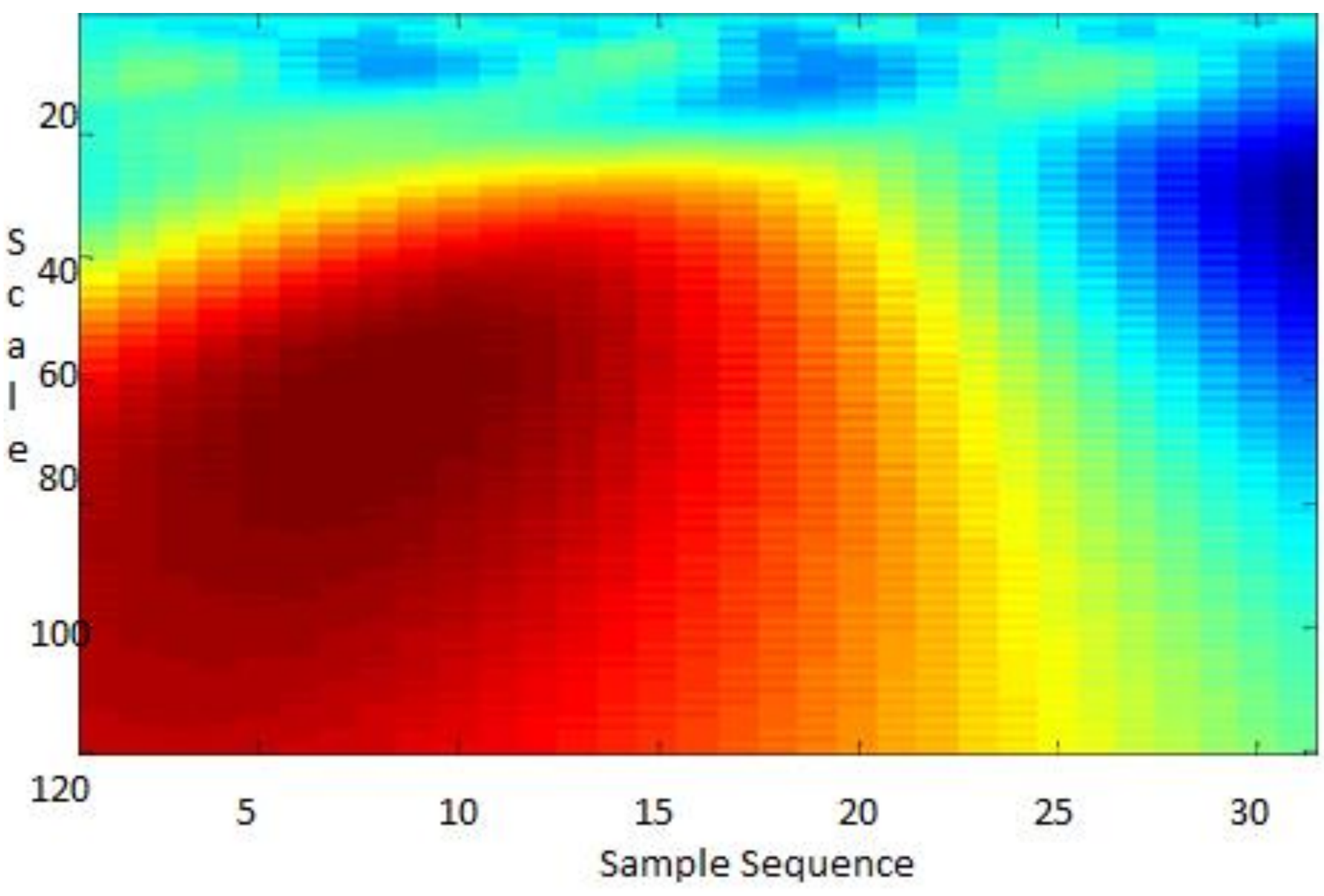

**Fig-2a** Daubechies-4 wavelet based approach for wind speed fluctuations (during a decade) plot in case of December

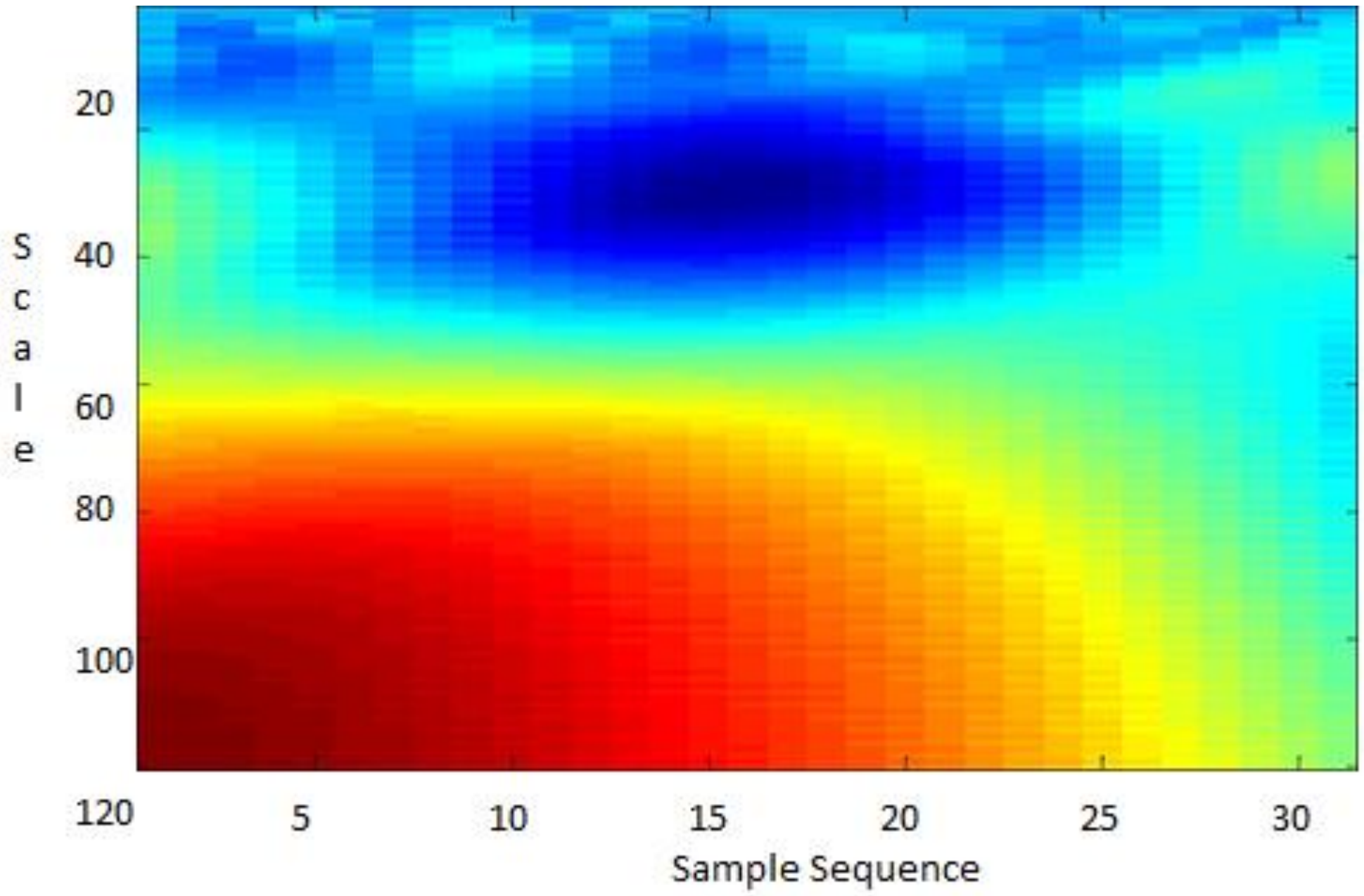

**Figure-2b** Daubechies-4 wavelet based approach for wind speed fluctuations (during a decade) plot in case of January

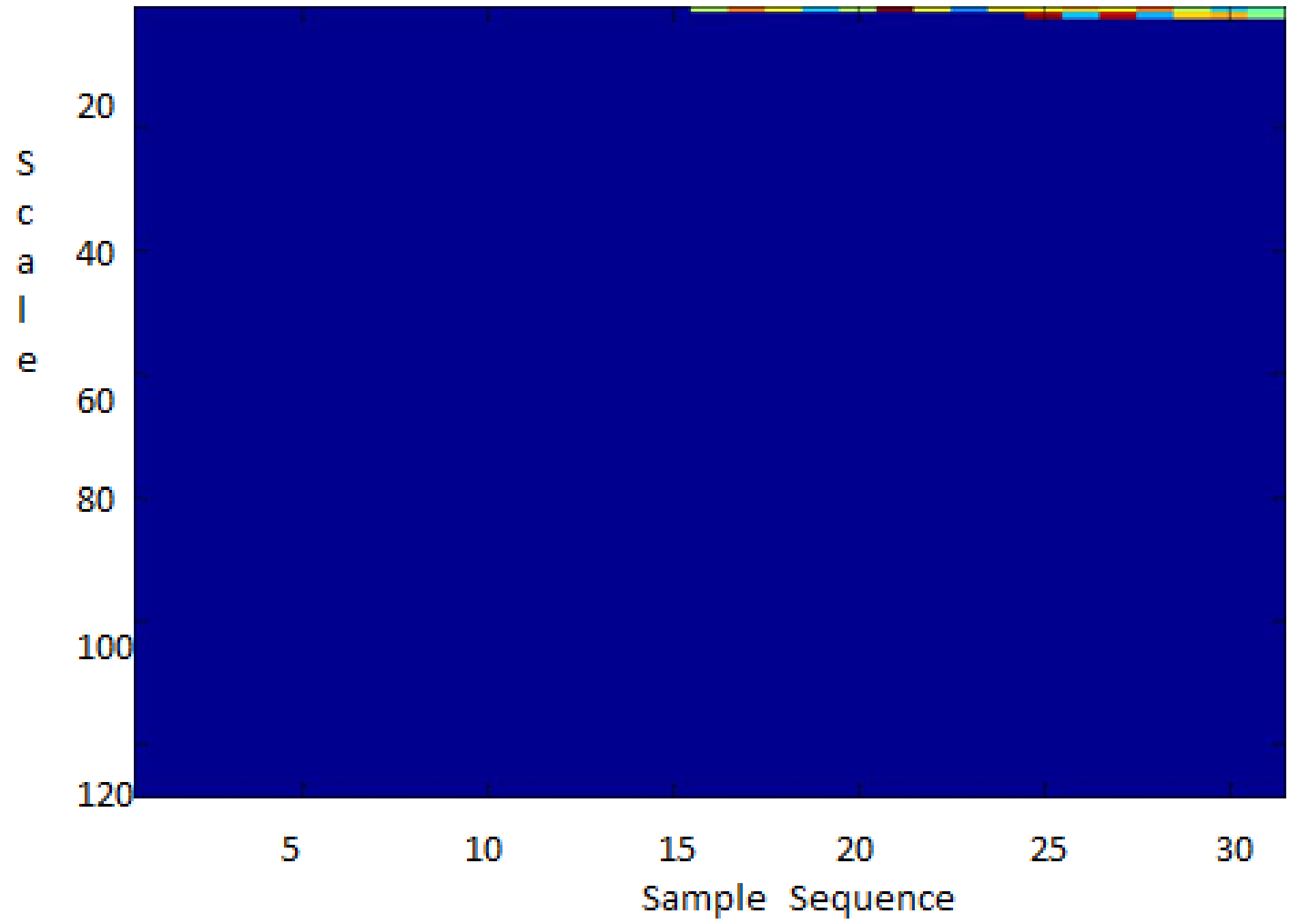

**Figure-2c** Daubechies-4 wavelet based approach for wind speed fluctuations (during a decade) plot in case of February

Hence it is clear that Db-4 is not efficient for keeping track of high wind speed fluctuation due to larger convection process.

Here it can be mentioned that other wavelets like symlets, coiflets, Morlets etc on highly agitated wind speed data were unable to get the desired results. Hence for resolving this kind of problem, S-transform is applied over median filtered wind speed fluctuation. The plots are given below Figure-3a to 3c.

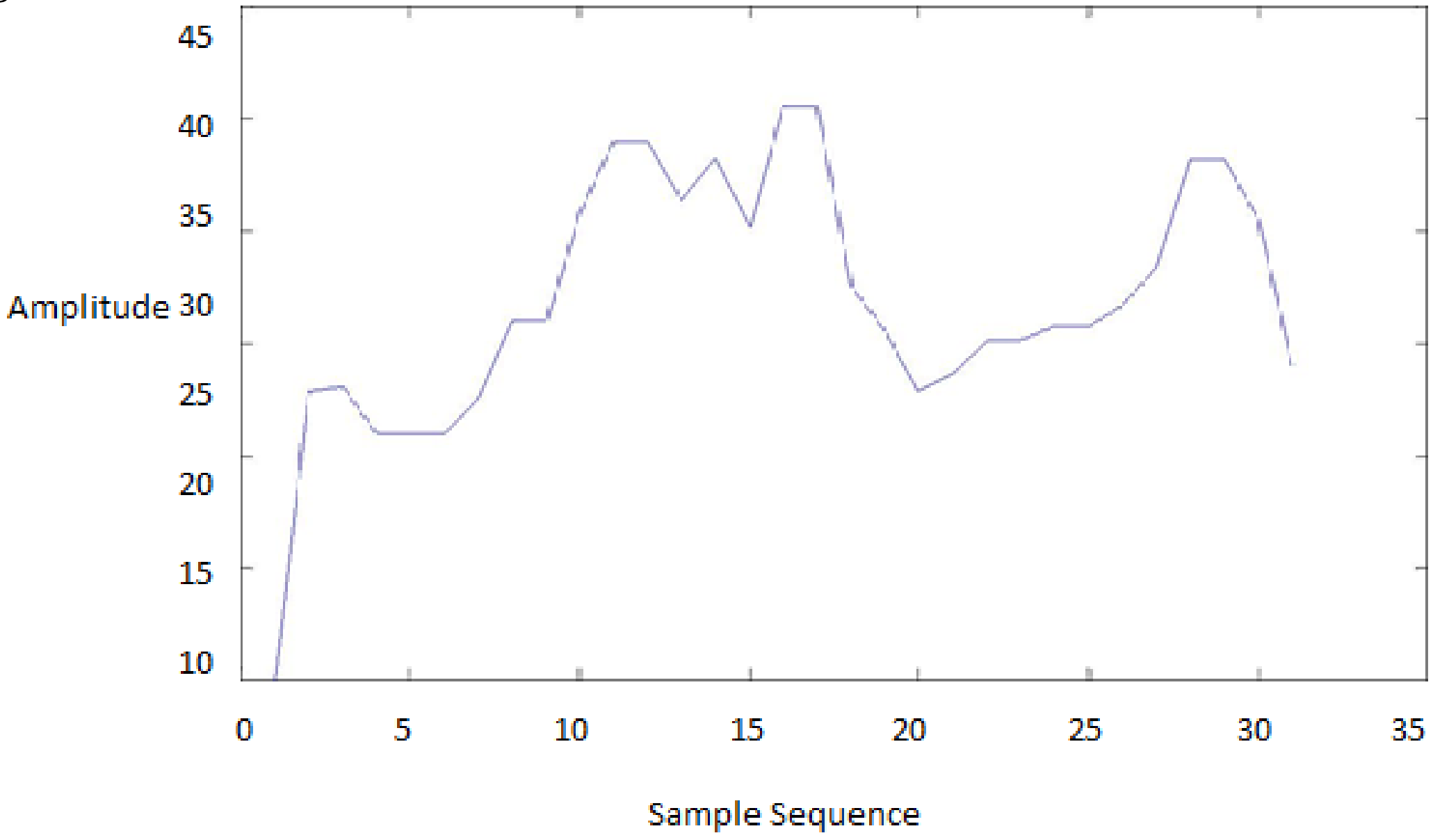

**Figure-3a** Average wind speed data (during a decade) plot of April after Median Filtering

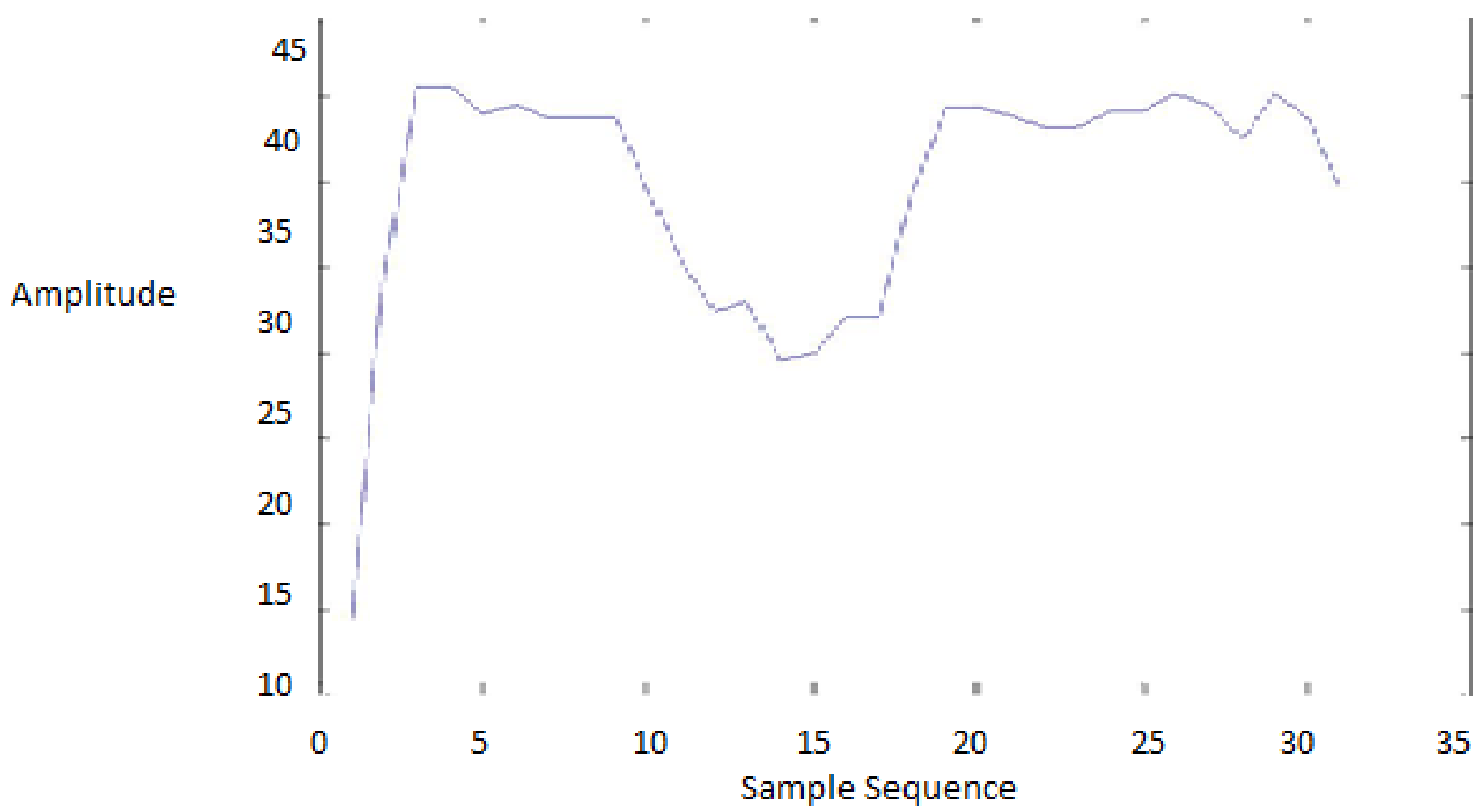

**Figure-3b** Average wind speed data (during a decade) plot of May after Median Filtering

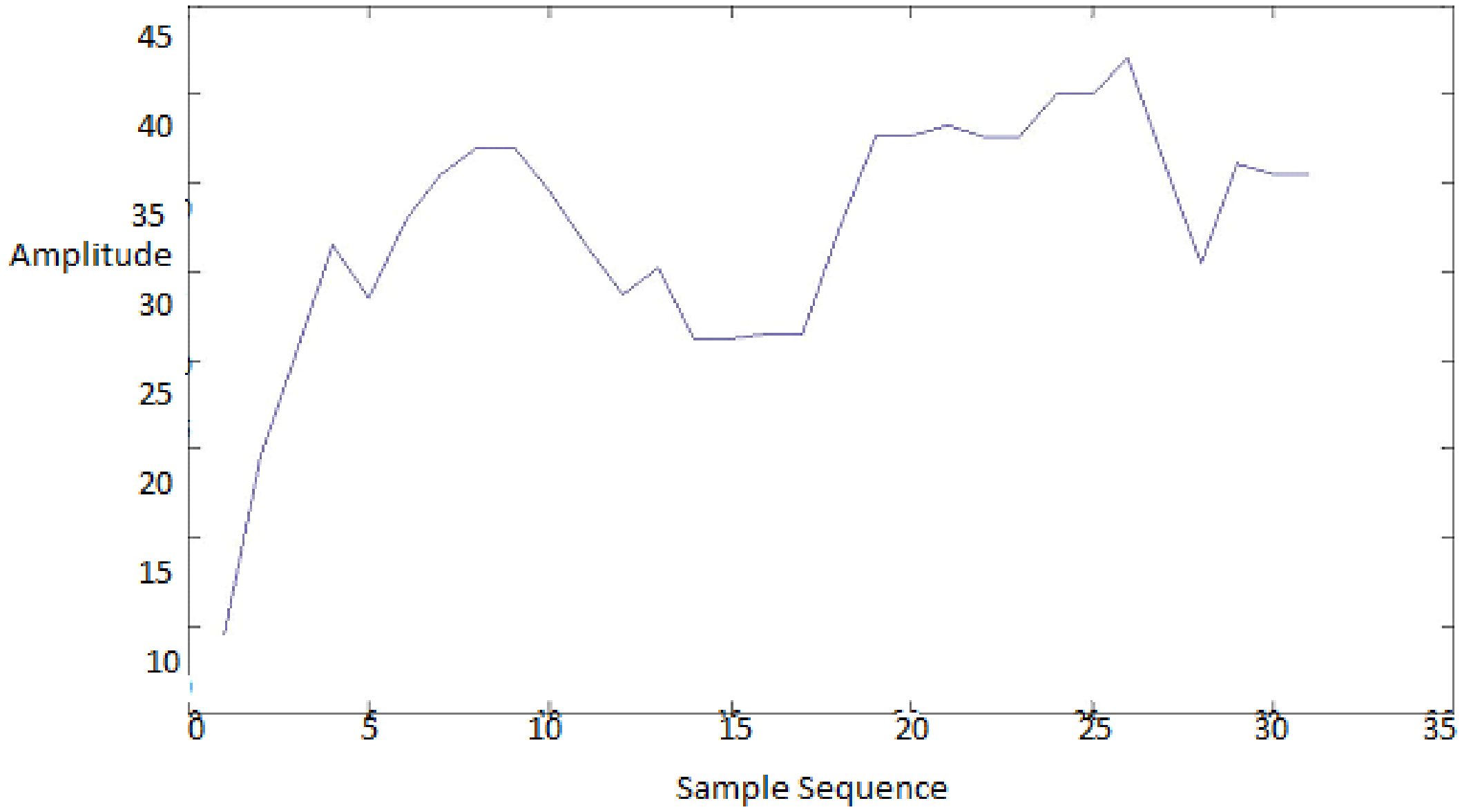

**Figure-3c** Average wind speed data (during a decade) plot of June after Median Filtering

After median filtering, wind speed fluctuation analysis is performed by S-transform to get the desired result. The plots are given in below Figure-4a to 4c.

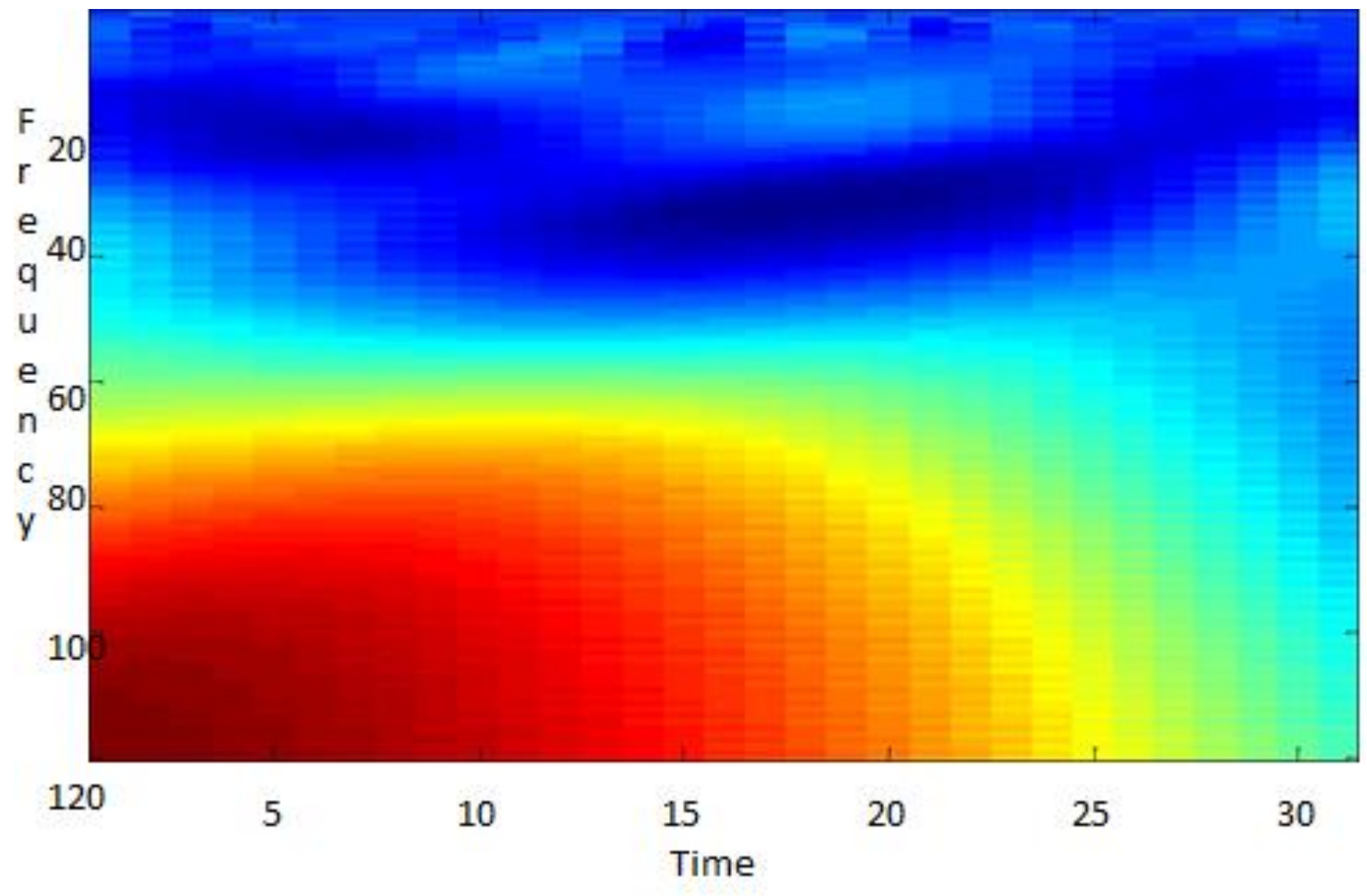

**Figure-4a** S-transform based approach for wind speed fluctuation (during a decade) in April

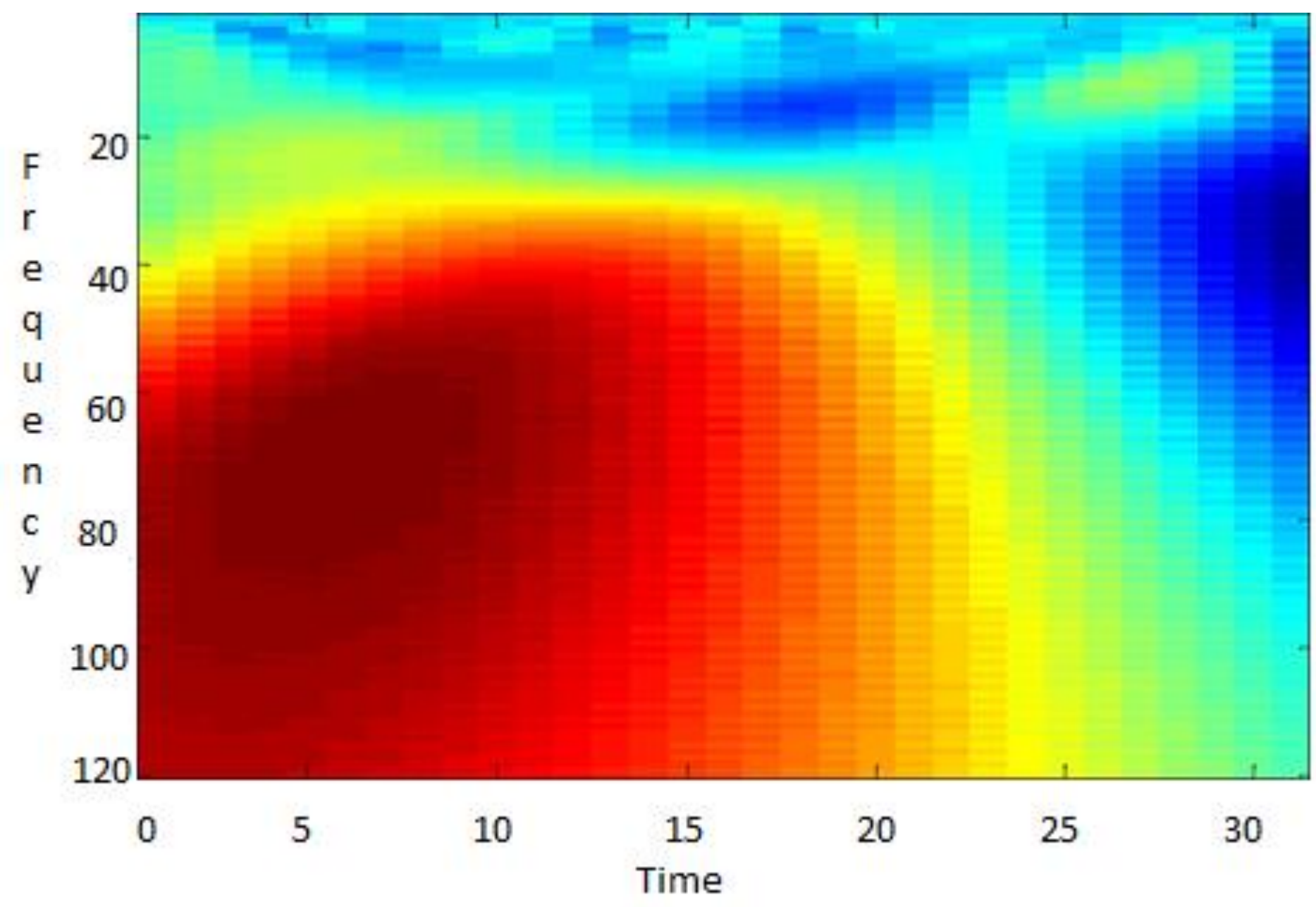

**Figure-4b** S-transform based approach for wind speed fluctuation (during a decade) in May

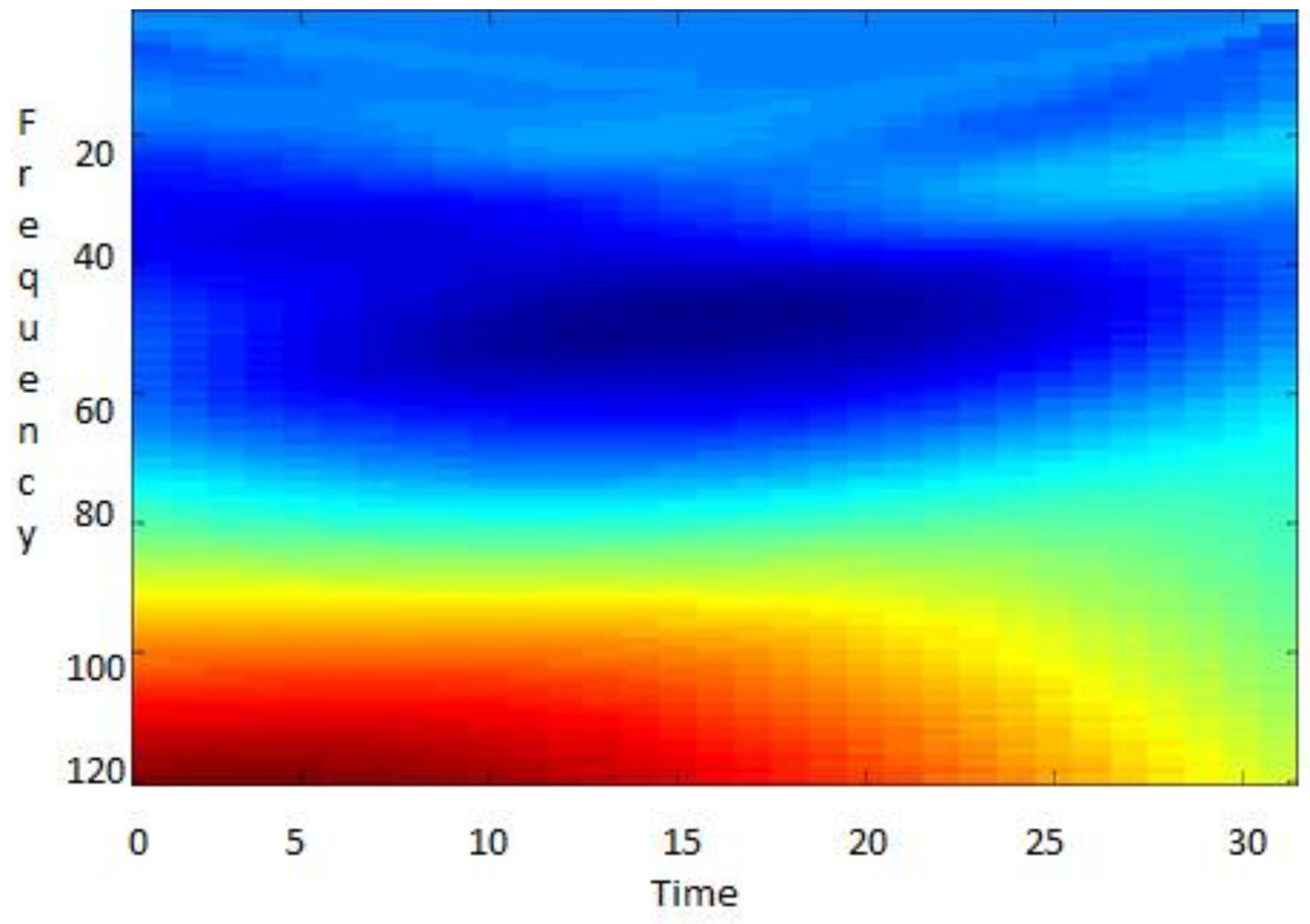

**Figure-4c** S-transform based approach for wind speed fluctuation (during a decade) in June

## 4. Conclusions

It is clear from the above results and discussions that Daubechies-4 wavelet is useful for extracting prominent features from winter wind data fluctuations. Due to large variation of wind fluctuation in summer, S-transform based approach is more appropriate.